\documentclass[aps,twocolumn,showpacs,amsmath,amssymb,prl,superscriptaddress]{revtex4}

\usepackage{graphicx}% Include figure files
\usepackage{dcolumn}% Align table columns on decimal point
\usepackage{bm}% bold math

\begin{document}

\title{Theory of Bubble Nucleation and Cooperativity in DNA Melting}

\author{Sa\'ul Ares} 
\affiliation{Theoretical Division and Center for Nonlinear Studies, 
Los Alamos National Laboratory, Los Alamos, New Mexico 87545, USA}
\affiliation{Grupo Interdisciplinar de Sistemas Complejos
(GISC) and
Departamento de Matem\'aticas,
Universidad Carlos III de Madrid, Avenida de la Universidad 30, 28911
Legan\'es, Madrid, Spain}
\author{N. K. Voulgarakis}
\affiliation{Department of Physics, University of Crete and Foundation for
Research and Technology-Hellas (FORTH), P.O. Box 2208, 71003 Heraklion, Crete, Greece}
\author{K. \O. Rasmussen}
\author{A. R. Bishop}
\affiliation{Theoretical Division and Center for Nonlinear Studies, 
Los Alamos National Laboratory, Los Alamos, New Mexico 87545, USA}

\date{\today}% It is always \today, today,

\begin{abstract}
The onset of intermediate states (denaturation bubbles) and their role during
the melting transition of DNA are studied using the Peyrard-Bishop-Daxuois model by
Monte Carlo simulations with no adjustable parameters.
Comparison is made with previously published experimental results finding
excellent agreement. Melting curves, critical DNA segment length for stability of bubbles and
the possibility of a two states transition are studied. 
\end{abstract}

\pacs{63.20.Pw,87.15.-v,87.15.He}
\maketitle
Accessing the genetic code stored in DNA is central to fundamental 
biological processes such as replication and transcription and this requires that 
the extraordinary stable double helical structure of the molecule must locally open to physically 
expose the bases. Although, in the cell, proteins may actively help separating the strands of 
double stranded DNA, recent evidence \cite{NAR,epl} corroborates that sequence-specific propensity to form strand
separations (bubbles) at transcription initiation sites exists and promotes thermal bubble formation.
Important thermal effects such as 
stability of different DNA sequences, and the properties of denaturation 
bubbles can be studied {\em in vitro} and provide important insight to the biological 
processes. Recent, experimental 
studies \cite{Zocchiepl,Zocchi,Zocchijmb} have
attempted to interrogate the nature and statistical significance of such bubble states. Intriguingly,
these experiments combine traditional UV absorption experiments with a novel bubble quenching technique that 
traps ensembles of bubbles to capture statistical properties of the bubbles.

The actual melting of double-stranded DNA occurs through an entropy driven phase transition.
The entropy gained in transitioning from the very rigid double-stranded DNA to the much more flexible single-stranded DNA
can, already at moderate temperatures, balance the energy cost of breaking a base-pair. Since, the double-stranded 
helix is held together by hydrogen bonds between complementary base-pairs: two bonds for the
AT pair and three bonds for the stronger GC pair, the sequence heterogeneity interplays with the 
entropy effects to create an extended premelting temperature window, (including the biologically
relevant regime) where large thermal bubbles are readily formed.  
Theoretical studies  of the melting transition have included ones based on Ising-type models \cite{Ising}
describing paired and unpaired bases, thermodynamics models like nearest-neighbor models \cite{NN}, Poland-Scheraga  models \cite{PS},
simple zipper models \cite{zipper,Ivanov}, or models that introduces a phenomelogical 
pairing potential between the bases \cite{PB,dpb,dp}. In particular the Peyrard-Bishop-Dauxois model \cite{dpb,dp}
is emerging as a model that is able to appropriately describe the melting transition 
but also the sequence dependence of the bubble nucleation {\it dynamics} in the pre-melting regime.

Here, we compare the powerful recent experimental results in Refs.
\cite{Zocchi,Zocchijmb} with Monte
Carlo simulations of the model proposed by Peyrard, Bishop, and Dauxois \cite{PB,dpb,dp}.
This model has already been successfully compared with denaturation experiments on short 
homogeneous sequences \cite{campa}.  The recent demonstration \cite{NAR} of the model's ability to accurately 
predict the locations at which large bubbles form in several viral sequences, is even more exceptional.
The difference
between our comparison and previous ones 
is that we use the same (deceptively) simple model, with no further
refinements that introduce new parameters that need to be fitted. Indeed 
parameters of the model are not changed to fit the experiments: we use the {\it same}
values for those parameters that were fixed in reference \cite{campa} for quite different
DNA sequences.

The potential energy of the model reads:

\begin{align}
\label{eq:Hamil}
V=&\sum_n \Big[D_n(e^{-a_ny_n}-1)^2 +\nonumber\\
&\frac{k}{2}(1+\rho e^{-\beta (y_n+y_{n-1})})(y_n-y_{n-1})^2 \Big]
\end{align}

The sum is over all the base-pairs of the molecule and $y_n$ denotes the relative
displacement from equilibrium at the $n^{th}$ base pair. The first term of the potential energy is a
Morse potential that represents the hydrogen bonds between the bases. The second
term is a next-neighbor coupling that represents the stacking interaction
between adjacent base pairs: it comprises a harmonic coupling multiplied by a term
that strengthens the coupling when the molecule is closed and makes it weaker
when it is melted, in this way taking into account the different stiffness (i.e. entropy effects) of DNA
double strands and single strands (this effect can be directly observed, in model calculations, in terms of a softning of the 
characteristic frequencies of the system with rising temperature \cite{nanoletter}).  This nonlinear coupling results in
long-range cooperative effects in the denaturation, leading to an abrupt
entropy-driven transition \cite{dpb,dp}. A crucial point for obtaining correct
results
is the accurate description of the heterogeneity of the sequence \cite{usuni}. In
this model it is incorporated by giving different values to the
parameters of the Morse potential, depending on the base-pair type of the site
considered: adenine-thymine (AT) or guanine-cytosine (GC). The parameter values
we have used are those used in Ref. \cite{campa}:  $k=0.025 eV/A^2$,
$\rho=2$, $\beta=0.35 A^{-1}$ for the inter-site coupling, while for the Morse
potential $D_{GC}=0.075 eV$, $a_{GC}=6.9 A^{-1}$ for a GC base pair, and
$D_{AT}=0.05 eV$, $a_{AT}=4.2 A^{-1}$ for an AT pair. These parameters were chosen to fit thermodynamic
properties of DNA \cite{campa}. One should be cautious in relating these parameters 
directly to microscopic properties, and recall that they arise as a result of several physical phenomena at the 
microscopic level. 

Using the standard Metropolis algorithm \cite{Metropolis,details}, we have performed Monte Carlo simulations
on this model \cite{comment}. For each temperature, we
performed a number of simulations. In each of these simulations, we compute the mean
profile $\langle y_n \rangle$,  from which we obtain the fraction of open base-pairs. We consider the 
n'th base-pair to be open if $\langle y_n \rangle$  exceeds a certain threshold. Applying the same threshold, we record at the 
end of each simulation whether the entire molecule was open (denaturated). 
Performing a large number of such simulations starting from different initial conditions we 
obtain the averaged fraction $f$ of open base-pairs and the averaged fraction of denaturated molecules $p$ at a given 
temperature.  In this way, we "simulate" the experiments,
where the measures are made over a large ensemble of
molecules.  The threshold we have used is 0.5A, but we have used other values and observed that the faction $p$ of 
denaturated molecules depends only very slightly on the
threshold value. The fraction of open base pairs, $f$, displays a somewhat stronger
dependence on the threshold value.
In the same manner as Ref. \cite{Zocchi} we obtain the 
averaged fractional length of the bubbles as $l=(f-p)/(1-p)$.
The experimental work \cite{Zocchi,Zocchijmb} concentrated on two sets of sequences one set
(\emph{bubble-in-the-middle} sequences) designed to form bubbles in the middle of the short 
sequence,  and another set (\emph{bubble-at-the-end} sequences) designed to form bubbles (openings) at one end 
of the sequences. Specifically, these sequences are:\\
(a) Bubble-in-the-middle  sequences :\\
L60B36:\ CCGCCAGCGGCGTTATTACATTTAATTC TTAAGTATTATAAGTAATATGGCCGCTGCGCC\\
L42B18:\ CCGCCAGCGGCGTTAATACTTAAGTATT ATGGCCGCTGCGCC\\
L33B9:\ CCGCCAGCGGCCTTTACTAAAGGCCGCT GCGCC\\
(b) Bubble-at-the-end sequences:\\
L48AS:\ CATAATACTTTATATTTAATTGGCGGCGC ACGGGACCCGTGCGCCGCC\\
L36AS:\ CATAATACTTTATATTGCCGCGCACGCGT GCGCGGC\\
L30AS:\ ATAAAATACTTATTGCCGCACGCGTGC GGC\\
L24AS:\ ATAATAAAATTGCCCGGTCCGGGC\\
%L19AS\_1:\ GCAGCGGCCTGGCCGCTGC\\
L19AS\_2:\ ATAATAAAGGCGGTCCGCC\\
%L13AS:\ GCCGCCAGGCGGC\\
%L11AS:\ CCGCCAGGCGG\\
\begin{figure}[h]
\vspace*{2mm}
\includegraphics[width=5.5cm]{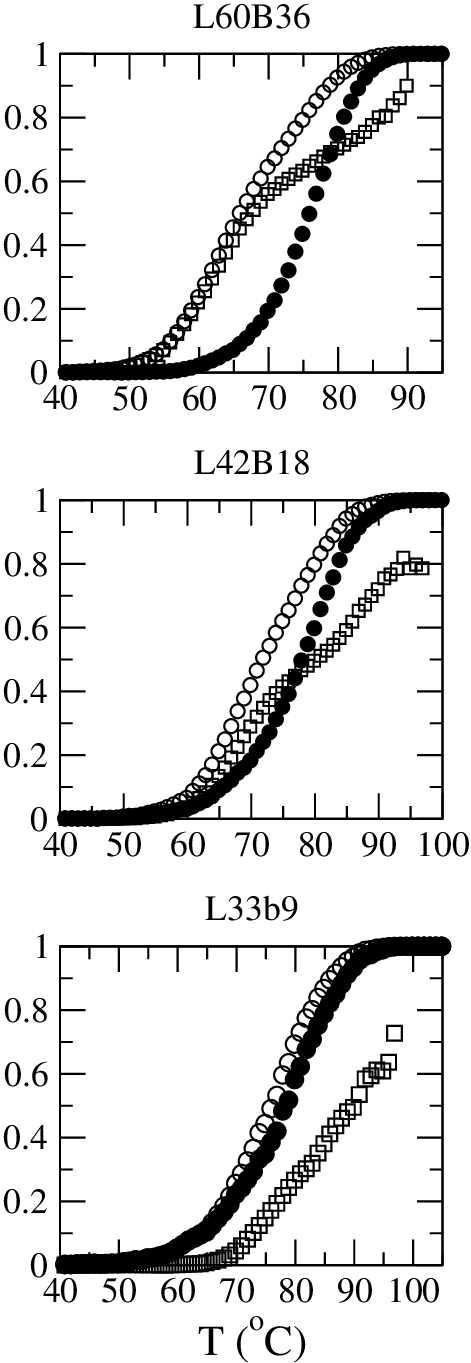}%\\[1mm]
\caption{\label{fig:k3}Melting profiles for the \emph{bubble-in-the-middle}
sequences \cite{Zocchiepl,Zocchi,Zocchijmb}. Filled circles are $p$, open circles are $f$ and squares are $l$.}
\end{figure}
The bubble-in-the-middle sequences are rich in AT base-pairs in the middle,
while the bubble-at-the-end are rich in AT base-pairs at one
end of the molecule. The AT base pairs are bonded by two hydrogen bonds, as opposed to
the stronger triple hydrogen bonding of the GC base-pairs. This fact is obviously reflected in the model parameters 
($D_{AT}=0.05$ while $D_{GC}=0.075$) and it also indicates that AT rich regions denaturate 
at lower temperatures that GC rich regions

%NOTE: AND L19AS\_1, L13AS AND L11AS? THEY ARE NOT FOR SURE BUBBLE IN THE END. ASK
%ZOCCHI. ALSO ASK HIM WHERE HE GETS FROM THE THREE POINTS FOR LOWER L IN
%FIG.\ 5 OF THE JMB PAPER.
\begin{figure}
\vspace*{2mm}
\includegraphics[width=7.0cm]{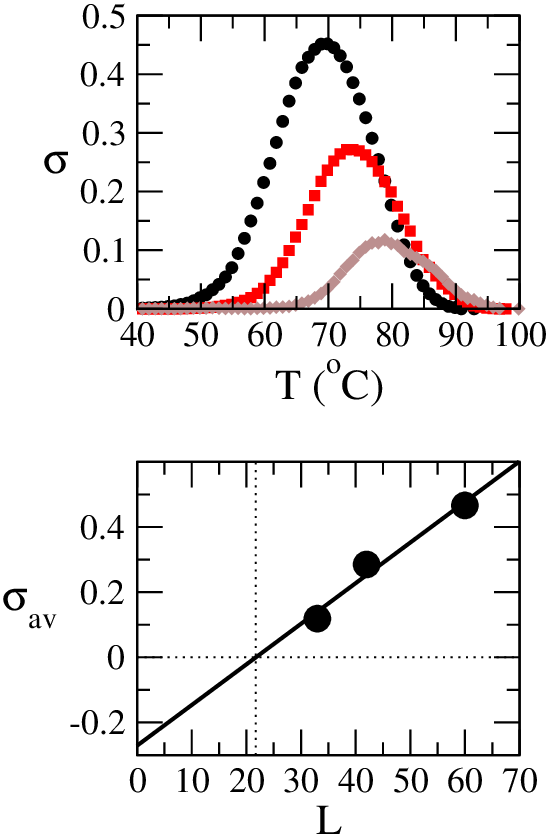}%\\[1mm]
\caption{\label{fig:s3t}Upper figure: $\sigma=f-p$ versus $T$
for L60B36 (circles), L42B18 (squares) and L33B9 (diamonds) \cite{Zocchiepl,Zocchi,Zocchijmb}.
Lower figure: $\sigma_{av}$
versus the length,  $L$, of the molecule.}
\end{figure}
\begin{figure}
\vspace*{2mm}
\includegraphics[width=5.5cm]{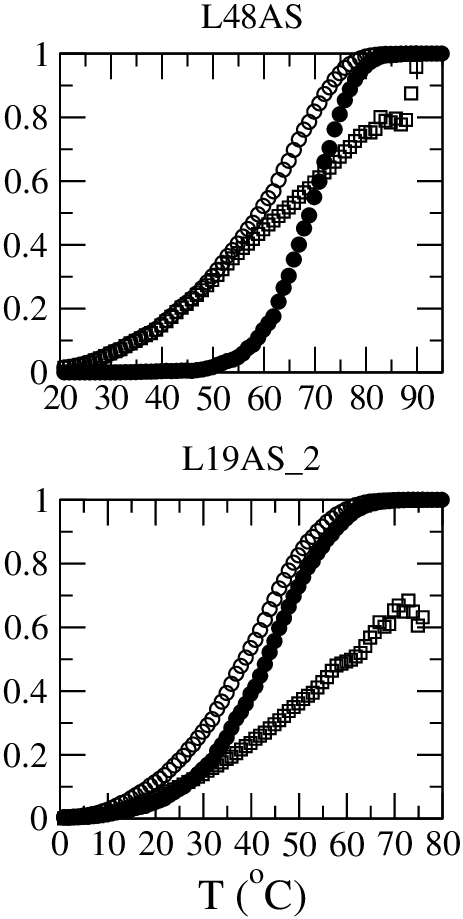}%\\[1mm]
\caption{\label{fig:l48} Melting profile for the L48AS and L19AS\_2 sequences.
 Symbols are as in Fig.\ \ref{fig:k3}.}
\end{figure}
\begin{figure}
\vspace*{2mm}
\includegraphics[width=7.0cm]{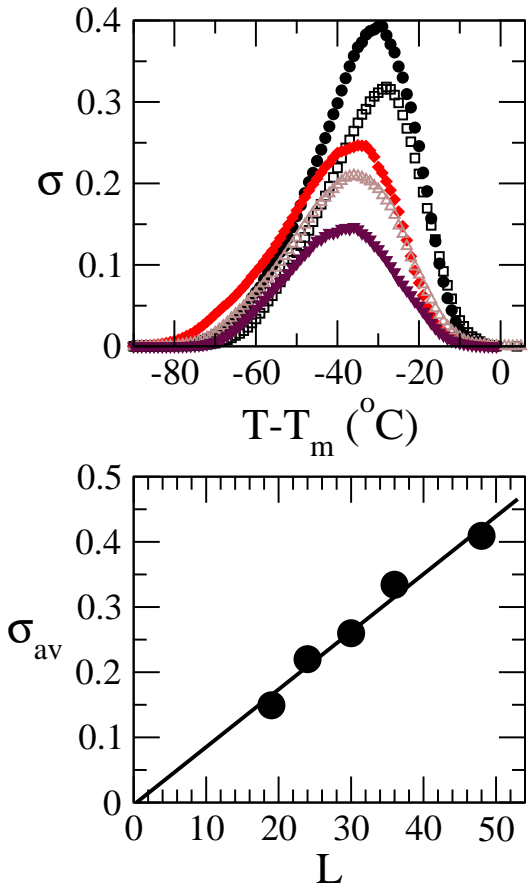}%\\[1mm]
\caption{\label{fig:s8t}Upper figure: $\sigma=f-p$ versus $T-T_m$ ($T_m$ is the
melting temperature) for L48AS (circles), L36AS (squares), L30AS (diamonds),
L24AS (triangles up) and L19AS\_2 (triangles down) \cite{Zocchiepl,Zocchi,Zocchijmb}. Lower figure: $\sigma_{av}$
versus $L$, the molecule's length.}
\end{figure}

In Fig.\ \ref{fig:k3} we present our results for the bubble-in-the-middle
sequences and we see a very good agreement with the
experimental results given in Refs. \cite{Zocchi,Zocchijmb}. As in the experimental results
we find for the L60B36 sequence that $f \approx l$ for $l < 0.6$. After this point, $l$ displays a 
plateau, resulting from the occurrence of completely denaturated molecules at $T \simeq 65 C$. As noted in the experimental 
work, the plateau occurs at $l \sim 0.6$ because this is the ratio
between the AT-rich central region, of
$36$ base pairs, and the molecule's total of $60$ base pairs.
The fact that $f \approx l$ before the plateau indicates that the
bubble opens continuously as a function of temperature until it reaches its full size, while there are 
very few completely melted molecules at these temperatures. For the L42B18 sequence we
again find a plateau in $l$ at the value $42/18 \approx 0.43$, but here
$f \neq l$ even at the lower temperatures (this is even more pronounced for L33B9). This shows 
that bubble generation and complete denaturation are both possible at lower temperatures.
Since the three sequences are similar in structure and merely differ in the length of AT-rich region, this demonstrates 
that for these structures bubble are only sustainable if the soft region is of size 20 base-pairs or more. 
To further illustrate this point we show in Fig.\ \ref{fig:s3t} $\sigma=f-p$, which
represents the fraction of bases participating in a bubble state at a given temperature. 
The upper figure of
Fig.\ \ref{fig:s3t} shows, as discussed, that as the soft bubble region becomes
shorter, bubble states become less important, as also seen
in Ref. \cite{Zocchi}.
In the lower figure, we summarize the length dependence of the
incidence of bubble states. We plot $\sigma_{av}$, the area under
the curves in the upper figure divided by their width, versus  the molecule
length, $L$. The line is a linear fit showing that these intermediate states
disappear ($\sigma_{av}=0$) for $L \approx 22$,
in excellent agreement with the experimental
conclusion in Refs. \cite{Zocchi,Zocchijmb}.

In Fig.\ \ref{fig:l48} we show the melting curves for two of the
bubble-in-the-end molecules. Comparison with experimental results in
ref. \cite{Zocchijmb} is again
good although not as good as in the the bubble-in-the-middle cases. This is due to the 
limitations of our model at the ends of the DNA molecule. 
Most experimental features are, however,  still reproduced.
For instance, in the L48AS sequence we find the same plateau on $l$ at
$L \approx 0.8$ that is seen in the experiments.
To overcome the problem caused by the boundaries, in Fig.\ \ref{fig:s8t} we
consider how $\sigma$ and $\sigma_{av}$ change with the system size, as the
deformation imposed by the excessive end opening will appear in all the
molecules and in that way will not contaminate the global picture. In the upper
figure we plot $\sigma$ versus $T-T_m$ ($T_m$ is the melting temperature),
finding that the bubble states are smaller for the shorter sequences, as in
Ref. \cite{Zocchijmb}. In the lower figure we plot $\sigma_{av}$ versus
$L$. The extrapolation to $\sigma_{av}=0$ occurs at a value compatible
with $L \approx 1$, as
in Ref. \cite{Zocchijmb}, and shows that in our model a two-state
transition for this kind of sequences
would only be possible in the limit $L \approx 1$, just as in the
experiments.

We have shown that the theoretical model proposed by Peyrard, Bishop, and Dauxois with no
further parameters or fitting, accurately reproduces experiments on DNA
denaturation, not only for the melting curve, but also for the formation and
role of bubble states in the premelting regime. Experimental
observations regarding nucleation size of the bubbles in the middle of a molecule and the
possibility of a two states transition are exactly recovered by the model. This
demonstrates that this model not only works for very
large DNA strands \cite{NAR}, but also for short strands such as the ones studied here.
Remarkably, these include both natural and synthetic structures.

We are grateful to Prof. G. Zocchi for insightful discussions of the data in Refs. \cite{Zocchiepl, Zocchi, Zocchijmb}. This work has been
supported in part by the Ministerio de Ciencia y Tecnolog\'\i a of Spain
through grant BFM2003-07749-C05-01 (SA). SA acknowledges financial
support from the Center for Nonlinear Studies, where this work was performed. Work at Los Alamos is performed
under the auspices of the US Department of Energy.


\begin{thebibliography}{88}
\bibitem{details} We use the standard Metropolis algorithm to produce an equilibrium state of the system: A single base pair, $n_0$, 
is picked at random and a new value of the variable $y_{n_0}$ is proposed according to a thermal (at the appropriate temperature)
 Gaussian distribution at this base pair. The proposed value is accepted according to the Metropolis 
probability: 1 if the energy $E_{new}$ of the new configuration is lower than that $E_{old}$ of the old configuration, and 
$\exp(-[E_{new}-E_{old}]/kT)$ if $E_{new}> E_{old}$. The process is continued after thermal equilibrium is reached in the measurement phase.
\bibitem{comment} We have used periodic boundary conditions for the bubbles-in-the-middle sequences and open boundary conditions for 
the bubbles-at-the-end sequences. While it may appear most reasonable to apply open boundary conditions, in both cases, we consistently find 
that such boundary conditions lead to a much too large propensity for openings at the ends of the DNA molecule. This represents a clear 
problem of the model that we are working towards improving.
\end{thebibliography}
\end{document}